\documentclass[12pt]{article}
\begin{document}
\newfont{\elevenmib}{cmmib10 scaled\magstep1}%
\renewcommand{\theequation}{\arabic{section}.\arabic{equation}}
\newcommand{\tabtopsp}[1]{\vbox{\vbox to#1{}\vbox to12pt{}}}
\font\larl=cmr10 at 24pt
\renewcommand{\thesection}{\Roman{section}}

\newcommand{\preprint}{
            \begin{flushleft}
   \elevenmib Yukawa\, Institute\, Kyoto\\
            \end{flushleft}\vspace{-1.3cm}
            \begin{flushright}\normalsize  \sf
            YITP-01-06\\
           {\tt hep-th/0102153} \\ February 2001
            \end{flushright}}
\newcommand{\Title}[1]{{\baselineskip=26pt \begin{center}
            \Large   \bf #1 \\ \ \\ \end{center}}}
\hspace*{2.13cm}%
\hspace*{0.7cm}%
\newcommand{\Author}{\begin{center}\large \bf
           R. Caseiro${}^a$, J.-P. Fran\c{c}oise${}^b$ and R.
Sasaki${}^c$ \end{center}}
\newcommand{\Address}{\begin{center}
            $^a$ Universidade de Coimbra,\ \
     Departamento de Matem\'atica\\
     3000 Coimbra, Portugal \\
     $^b$ Universit\'e de Paris 6, Laboratoire ``GSIB"\\
     125 Rue du Chevaleret, 75013, Paris, France\\
     ${}^c$ Yukawa Institute for Theoretical Physics\\
     Kyoto University, Kyoto 606-8502, Japan
      \end{center}}
\newcommand{\Accepted}[1]{\begin{center}{\large \sf #1}\\
            \vspace{1mm}{\small \sf Accepted for Publication}
            \end{center}}
\baselineskip=20pt

\preprint
\thispagestyle{empty}
\bigskip
\bigskip
\bigskip

\Title{Quadratic Algebra associated with Rational Calogero-Moser Models}
\Author

\Address
\vspace{1cm}

\begin{abstract}
Classical Calogero-Moser models with rational potential are known to be
superintegrable. That is, on top of the
$r$ involutive conserved quantities necessary for the integrability of
a system with $r$  degrees of freedom, they possess an additional
set of  $r-1$ algebraically and functionally
independent globally defined conserved quantities.
At the quantum level, Kuznetsov uncovered the existence of
a quadratic algebra structure as an
underlying key for superintegrability for the models based on
$A$ type root systems. Here we demonstrate in a universal way
the quadratic algebra structure
for quantum rational Calogero-Moser models based on any root systems.
\end{abstract}
\bigskip
\bigskip
\bigskip

%%%%%%%%%%%%%%%
\section{Introduction}
\label{intro}
\setcounter{equation}{0}

Calogero-Moser models \cite{Cal, Sut},
\cite{CalMo} with the rational potentials, without the
harmonic confining force,
have the simplest and  best understood dynamical structures among
models with the other types of potentials.
Their superintegrability, that is the existence of
$2r-1$ global, functionally independent conserved quantities
(constants of motion) for a system of $r$ degrees of freedom,
is one of the most striking features. It was found  at the classical
level by Wojciechowski
\cite{Woj} and at the quantum level by Kuznetsov \cite{Kuz}
and Ujino-Wadati-Hikami \cite{UjWa},
for models based on the $A$ type root systems.
Kuznetsov \cite{Kuz} uncovered an interesting algebraic structure,
the so-called {\em quadratic algebra\/}
as a hidden symmetry of the superintegrability.
Ujino-Wadati-Hikami \cite{UjWa} introduced a similar algebraic structure.
The concept of superintegrability is closely related with that of
{\em algebraic linearizability\/} formulated by Caseiro-Fran\c{c}oise
\cite{cf} and developed further by Caseiro-Fran\c{c}oise-Sasaki
\cite{cfs} for the models based on any root systems.
We follow the notation of our previous paper unless otherwise stated.

%\bigskip
In this paper we show, at the quantum level,
that the quadratic algebra is `universal', namely,  it is
enjoyed by all the  rational Calogero-Moser models based on {\em any}
root systems including the non-crystallographic ones.
The same assertion at the classical level simply follows as the
classical limit of replacing the quantum commutator by the Poisson
bracket.
The generators of the quadratic algebra are
 the  above mentioned conserved quantities of the
superintegrable theory. Among them, the involutive subset of $r$
conserved quantities, which characterize the Liouville integrability
of the system with $r$ degrees of freedom, constitute the Cartan
subalgebra and an ideal among the conserved quantities.
Commutators among the additional conserved quantities turn out to be
bi-linear
(quadratic)
 combinations of the two types of conserved quantities.
 This non-linear algebra seems to be closely related
to
the $W$-algebras \cite{Zam}, extensions of the Virasoro algebra,
or to the algebras related with  the $R$-matrices of integrable theories
\cite{Freidel-Maill} but
the precise relationship remains to be clarified.

Calogero-Moser models for any root systems were formulated by
Olshanetsky and Perelomov \cite{OP1}, who  provided Lax pairs
for the models based on the classical root systems, {\em i.e.\/}
the $A$, $B$, $C$, $D$ and $BC$ type root systems.
A universal {\em classical\/} Lax pair applicable to all
the Calogero-Moser models
based on any root systems including the $E_{8}$ and the
non-crystallographic root systems was derived by Bordner-Corrigan-Sasaki
\cite{bcs2}
which unified various types of Lax pairs known at
that time
\cite{DHoker_Phong, bcs1}.
A universal {\em quantum\/} Lax pair applicable to all the Calogero-Moser
models
based on any root systems and for degenerate potentials
 was derived by Bordner-Manton-Sasaki \cite{bms}
 which provided the basic tools
for the present paper.

The purpose of the present paper is twofold.
Firstly, to derive and present the
{\em quadratic algebra\/} for rational  Calogero-Moser models based on
any root systems in its fullest universality based on the universal
Lax pair \cite{bms}. Extracting detailed information from the
quadratic algebra to elucidate dynamical properties of each specific
system would require  formulations suitable for the
particular systems. This would not be discussed here. Secondly, we
formulate and present the quantum versions of
 various concepts and formulas
related to the algebraic linearizability introduced and
developed in
\cite{cfs}. As has been noticed from the earliest days of
Calogero-Moser models, the quantum and classical integrability
are very closely related. Many formulas related to  the
algebraic linearizability take the same form at the classical
and quantum levels, with some notable exceptions as will be
mentioned in the paper.

This paper is organized as follows. In section two we introduce
the model and notations with an emphasis on the difference
between the quantum and classical versions. The quantum theorem
of the algebraic linearizability for the rational model is derived
based on the Lax pair formalism.
In section three we evaluate fundamental commutation relations
which are necessary for the quadratic algebra.
This will be carried out with the help of the Dunkl operators, or the
so-called $\ell$ operators which are equivalent to the quantum $L$
operator.
The quantum theorem of the algebraic linearizability for the
higher Hamiltonians of
the rational model is derived.
In section four the quadratic algebra for rational Calogero-Moser models
is derived and presented in its fullest universality.
Section five gives the quantum version of the algebraic linearizability
of the rational potential model with harmonic confining force.
The problem of quantum integrability of rational Calogero-Moser model
with quartic interactions is not yet settled.
In section six we present a partial result
that the quantum equations of motion can be cast into
 Lax type matrix equations. The existence of quantum conserved
quantities, however, does not follow from these matrix
equations. In section seven the quantum version of the
algebraic linearizability for trigonometric (hyperbolic)
Calogero-Sutherland models is given for those models based on
root systems which have minimal representations. The final
section is for comments on the hermiticity of the algebra
generators.

 %%%%%%%%%%%%%%%%%%%%%%%%%%%%%%%%%%%%%%%%%%%%%
\section{Quantum Calogero-Moser Models with Rational Potential}
\label{cal-mo}
\setcounter{equation}{0}

Let us start with the Hamiltonian of {\em quantum} Calogero-Moser model
with rational potential based on any root system, which could be any
one of the crystallographic root systems, $A_{r}$, $B_{r}$, $C_{r}$,
$D_{r}$, ($BC_{r}$), $E_{6}$, $E_{7}$, $E_{8}$, $F_{4}$ and $G_{2}$
or the non-crystallographic $H_{3}$, $H_{4}$ and $I_2(m)$, which is the
dihedral root system associated with a regular $m$-gon.
The existing works on the quadratic algebras are all for the $A_{r}$
root system \cite{Kuz, UjWa, Gonera1}.
Let us denote by $\Delta$ a root system of rank $r$. The
dynamical variables are  the coordinates $q_{i},
i=1,...,r$ and their canonically conjugate momenta $p_{i},
i=1,...,r$,  with
the canonical commutation relations:
\begin{equation}
   [q_{j},p_{k}]=i\delta_{jk},\qquad [q_{j},q_{k}]=
   [p_{j},p_{k}]=0,\quad j,k=1,\ldots,r.
\end{equation}
As usual the momentum operator \(p_j\) acts as a derivative operator
on a (wave) function $f$ of $q$:
\[
  f\to p_jf: \quad (p_jf)(q)=-i{\partial f(q)\over{\partial q_j}},
  \quad j=1,\ldots,r.
\]
The Hamiltonian for the {\em quantum} Calogero-Moser model  with
rational potential is very simple:
\begin{equation}
    {\cal H}={1\over2}p^2+
    {1\over2}\sum_{\rho\in\Delta_+}
   {g_{|\rho|}(g_{|\rho|}-1)|\rho|^{2}\over{(\rho\cdot q)^2}},
   \quad \Delta_+: \mbox{set of positive roots},
   \label{ratHam}
\end{equation}
in which the real coupling constants \(g_{|\rho|}>0\) are
defined on orbits of the corresponding
finite reflection  group, {\it i.e.} they are
identical for roots in the same orbit.
The only difference with the classical Hamiltonian is the
coupling constant dependence, $g_{|\rho|}(g_{|\rho|}-1)$ instead of
$g_{|\rho|}^2$ in the classical case.
The Hamiltonian is invariant under reflections of the phase space
variables about a hyperplane perpendicular to any root
\begin{equation}
  {\cal H}(s_{\alpha}(p),s_{\alpha}(q))={\cal H}(p,q), \quad
   \forall\alpha\in\Delta,
  \label{HamCoxinv}
\end{equation}
with the action of \(s_{\alpha}\) on any vector
\(\gamma\in{\bf R}^r\) defined by
\begin{equation}
   \label{Root_reflection}
   s_{\alpha}(\gamma)=\gamma-(\alpha^{\vee}\!\!\cdot\gamma)\alpha,\quad
   \alpha^{\vee}\equiv2\alpha/|\alpha|^{2}.
\end{equation}

The integrability is best understood in terms of the quantum Lax pair
\cite{bms} or
the Dunkl operators \cite{Dunk,Heck}, which are known to be equivalent
with the Lax operator \cite{kps}.
Let us choose a set of \(\mathbf{R}^{r}\)
vectors
\({\cal R}=\{\mu^{(k)}\in\mathbf{R}^{r},\ k=1,\ldots, D\}\),
which form a \(D\)-dimensional representation of the Coxeter group.
That is, they are permuted
under the action of the Coxeter group and they form a single orbit.
For example, we can choose the
set of vector (minimal) weights for $A_{r}$ or $D_{r}$, or the set of
short (long) roots for $B_{r}$, $C_{r}$ or $F_{4}$, $G_{2}$ or the set
of all roots for $E_{6}$ to $E_{8}$.
Then the
Lax operators are \(D\times D\) dimensional matrices
\begin{eqnarray}
   \label{LaxOpDef}
   L(p,q) &=& p\cdot\hat{H}+X(q),\qquad X(q)
   =i\sum_{\rho\in\Delta_+}g_{|\rho|}
\frac{\rho.\hat{H}}{\rho\cdot q}\hat{s_\rho},
   \\ \nonumber
   M(q) &=&
   -\frac{i}{2}\sum_{\rho\in\Delta_+}g_{|\rho|}
   \frac{|\rho|^{2}}{(\rho.q)^2}(\hat{s_\rho}-I),
\end{eqnarray}
consisting of operators \(\{\hat{H}_j\}\),
(\(j=1,\ldots,r\)), \(\{\hat{s}_{\rho}\}\) and
the identity operator $I$.
Their matrix elements are defined by:
\begin{equation}
    (\hat{H}_{j})_{\mu\nu}=\mu_j\delta_{\mu\nu},\quad
    (\hat{s}_{\rho})_{\mu\nu}=\delta_{\mu,s_\rho(\nu)}=
      \delta_{\nu,s_\rho(\mu)},\quad \mu,\nu\in{\cal R}.
\end{equation}
The Lax operators are Coxeter covariant:
\begin{equation}
L(s_{\alpha}(p),s_{\alpha}(q))=\hat{s}_\alpha L(p,q)\hat{s}_\alpha,\quad
M(s_{\alpha}(q))=\hat{s}_\alpha M(q)\hat{s}_\alpha,
\label{lmcoxcov}
\end{equation}
and $L$ is hermitian $L^\dagger=L$ and $M$
is anti-hermitian $M^\dagger=-M$.

We see that the Heisenberg equations of motion are equivalent to a
matrix equation \cite{bms,kps}
\begin{equation}
  {dL\over{dt}}\equiv\dot{L}=i[{\mathcal H},L]=[L,M],
\end{equation}
in which the matrix elements are quantum operators. This means that
in general the {\em trace} of the product of
two matrix operators $A(p,q)$ and
$B(p,q)$  is not commutative, Tr$AB\neq$ Tr$BA$, or Tr$[A,B]\neq0$,
implying that Tr$L^{n}$ is not {\em conserved} in quantum theory.
However, thanks to the special property of the above $M$ matrix
\begin{equation}
    \sum_{\mu\in\mathcal{R}}M_{\mu\nu}
=\sum_{\nu\in\mathcal{R}}M_{\mu\nu}=0,
    \label{zerosum}
\end{equation}
the {\em total sum} of the powers of the Lax operator $L$ defined by
\begin{equation}
   F_{j}= \mbox{Ts}(L^j)\equiv\sum_{\nu,\mu\in{\mathcal
   R}}(L^j)_{\mu\nu},
   \quad j=0,1,\ldots, D-1,
    \label{jtotsum}
\end{equation}
is conserved:
\begin{eqnarray*}
 \frac{d}{dt}\mbox{Ts}(L^j) &= &\sum_{\mu,\nu\in{\mathcal
  R}}\left[(L^jM)_{\mu\nu}-(ML^j)_{\mu\nu}\right]\\
  &=& \sum_{\mu,\nu, \kappa\in{\mathcal R}}\left[
  L^j_{\mu\kappa}M_{\kappa\nu}-M_{\mu\kappa}L^j_{\kappa\nu}\right]\\
  &=& \sum_{\mu,\kappa\in{\mathcal R}}L^j_{\mu\kappa}
(\sum_{\nu\in{\mathcal
  R}}M_{\kappa\nu})- \sum_{\nu,\kappa\in{\mathcal R}}
(\sum_{\mu\in{\mathcal
  R}}M_{\mu\kappa})L^j_{\kappa\nu}=0.
\end{eqnarray*}
It is easy to see  from (\ref{lmcoxcov}) that $\{F_j\}$'s
are Coxeter invariant.
These form the involutive set of conserved quantities of the theory.
Not all of them are independent. As is well-known, the independent
conserved quantities appear for such $j$  as 1 plus {\em exponent}
of the root system, (see, for example, \cite{kps,ks2}).
For some choice of the set of vectors ${\cal R}$ for some root system
$\Delta$,
$F_j$ can be vanishing for certain $j$'s.
For example, if ${\cal R}$ contains a vector $\mu$ and $-\mu$ at the same
time
then $F_{odd}\equiv0$.

The Hamiltonian (\ref{ratHam}) is proportional to $F_{2}$,
\begin{equation}
 {\cal H}={1\over{2C_{\cal R}}} F_{2}
={1\over{2C_{\cal R}}}\mbox{Ts}(L^2),
\end{equation}
in which the coefficient $C_{\cal R}$ is defined by
\begin{equation}
   \mbox{Ts}(\hat{H}_j\hat{H}_k)
=\sum_{\mu\in{\cal R}}\mu_j\mu_k=\delta_{jk}
C_{\cal R}.
\label{defcr}
\end{equation}

Following the line of argument of \cite{cfs} we define
\begin{equation}
    Q=q\cdot\hat{H},\quad G_{j}=\mbox{Ts}(QL^j),
    \quad G_{j}^{(2)}=\mbox{Ts}(Q^{2}L^j),\quad j=0,1,\ldots, D-1,
    \label{Qdef}
\end{equation}
in which the last quantity $Q^{2}L^j$ was introduced by Ra\~nada
\cite{Ranada}.
Under the reflection, $Q$ transforms in the same way as $L$ and $M$,
(\ref{lmcoxcov}):
\begin{equation}
  q\to s_{\alpha}(q), \quad
Q(s_{\alpha}(q))= \hat{s}_\alpha Q(q)\hat{s}_\alpha.
\label{qcoxcov}
\end{equation}
Thus $G_j$ and $G_{j}^{(2)}$ are Coxeter invariant, too.
The time evolution of $Q$ is exactly the same as in the classical
case \cite{cfs}
\begin{equation}\label{RQ}
  \dot{Q}=[Q,M]+L,
\end{equation}
leading to the corresponding result:
\begin{eqnarray}
    \dot{G_j} &=& \mbox{Ts}(\dot{Q}L^j+Q\dot{L^j})\nonumber\\
  &=& \mbox{Ts}(QML^j-MQL^j+L^{j+1}+QL^jM-QML^j)\nonumber\\
  &=& \mbox{Ts}(L^{j+1})-\sum_{\nu,\kappa\in{\mathcal
  R}}(\sum_{\mu\in{\mathcal R}}M_{\mu\kappa})(QL^j)_{\kappa\nu}+
  \sum_{\mu,\kappa\in{\mathcal
  R}}(QL^j)_{\mu\kappa}(\sum_{\nu\in{\mathcal
  R}}M_{\kappa\nu})\nonumber\\
  &=& \mbox{Ts}(L^{j+1})=F_{j+1}.
    \label{Fgrel}
\end{eqnarray}
Like $\{F_{j}\}$'s not all of $\{G_{j}\}$'s are independent.
Independent $\{G_{j}\}$ appear when $\{j\}$
are the exponents of $\Delta$.
This provides the algebraic linearization of the quantum models.
Like in the classical theory we have:

\bigskip
{\bf Proposition \ref{cal-mo}.1}

The quantum Calogero-Moser system (\ref{ratHam}) is superintegrable for
any root system.

{\bf Proof.}
        On top of  the $D$ first integrals $F_{k}$ which are in involution,
we have
the $D(D-1)/2$ extra first integrals defined by
\begin{eqnarray}
H_{k,k'}&=&F_{k+1}G_{k'}-F_{k'+1}G_{k},\label{defHjk}\\
\dot{H}_{k,k'}&=&i[{\cal H}, H_{k,k'}]=0.
\label{Hjkcons}
\end{eqnarray}
Like in our previous paper for the classical systems \cite{cfs},
we do not demonstrate
that these $D(D-1)/2$ $\{H_{k,k'}\}$'s contain $r-1$
algebraically independent ones. That would require detailed exhaustive
arguments for each root system.
We refer to \cite{kps} for general arguments of independence of $\{F_j\}$
type conserved quantities.

\noindent
For the quantum models based on the $A$ type root system,
a similar result was derived by Gonera \cite{Gonera1} based
  on a $sl(2,{\bf R})$ representation.
The time evolution of $G_{j}^{(2)}$ is slightly complicated:
\begin{eqnarray}
    \dot{G}_{j}^{(2)} & = &
    \mbox{Ts}(\dot{Q}QL^{j})+\mbox{Ts}(Q\dot{Q}L^{j})
     +\mbox{Ts}(Q^{2}\dot{L}^{j})
    \nonumber  \\
     & = & \mbox{Ts}(LQL^{j})+\mbox{Ts}(QL^{j+1}).
    \label{Hevo}
\end{eqnarray}
Since $L$ and $Q$ do not commute in quantum theory, the classical
relation \(\dot{G}_{j}^{(2)}=2\mbox{Tr}(QL^{j+1})=2G_{j+1}\) does not
hold any longer. In quantum theory we have
\begin{equation}
    QL-LQ=i\delta_{kl}\hat{H}_{k}\hat{H}_{l}+iK,
    \quad
    K\equiv
\sum_{\rho\in\Delta_{+}}g_{|\rho|}(\rho\cdot\hat{H})
    ({\rho}^{\vee}\!\cdot\hat{H})\hat{s}_{\rho}.
    \label{QLcomm}
\end{equation}
The right hand side gives the `quantum corrections'. Thus we arrive at
\begin{equation}
    \dot{G}_{j}^{(2)}=2\mbox{Ts}(QL^{j+1})-i\delta_{kl}\mbox{Ts}
    (\hat{H}_{k}\hat{H}_{l}L^j)-i\mbox{Ts}(KL^j).
    \label{hatHeq}
\end{equation}
The second term is easy to evaluate, since
\begin{equation}
    \delta_{kl}\mbox{Ts}
    (\hat{H}_{k}\hat{H}_{l}L^j)=\delta_{kl}\sum_{\mu,\nu\in{\cal
    R}}(\mu_{k}\mu_{l}(L^j)_{\mu\nu})=\mu^2\mbox{Ts}(L^j),
    \label{qcfirst}
\end{equation}
in which $\mu^2$ is the same  for all $\mu\in{\cal R}$. The third
term reads
\begin{displaymath}
\mbox{Ts}(KL^j)=\sum_{\rho\in\Delta_{+}}\sum_{\mu,\nu\in{\cal
R}} g_{|\rho|}(\rho\cdot\mu)
    ({\rho}^{\vee}\!\cdot\mu)(L^{j})_{\mu\nu},
\end{displaymath}
and for any vector $\mu\in{\bf R}^{r}$ we have
\begin{equation}
    \sum_{\rho\in\Delta_{+}}g_{|\rho|}(\rho\cdot\mu)
    ({\rho}^{\vee}\!\cdot\mu)=
    {2\over{r}}\mu^{2} \sum_{\rho\in\Delta_{+}}g_{|\rho|},
    \label{Keig}
\end{equation}
in which
\begin{equation}
     {2\over{r}} \sum_{\rho\in\Delta_{+}}g_{|\rho|}
    \label{defCox}
\end{equation}
can be considered as a {\em deformed Coxeter number}. For
\(g_{|\rho|}\equiv1\)  it reduces to the Coxeter number. Thus we
arrive at a quantum formula
\begin{eqnarray}
    \dot{G}_{j}^{(2)} & = & 2G_{j+1}-i\mu^{2}
    \left(1+{2\over{r}} \sum_{\rho\in\Delta_{+}}g_{|\rho|}\right)F_{j}
    \nonumber  \\
     & = &  2G_{j+1}-i\mu^{2}{2\over{r}}\tilde{\cal E}_{0}F_{j}.
    \label{Hjquant}
\end{eqnarray}
Here,  the coefficient of the quantum corrections term
\(\tilde{\cal E}_{0}\) is defined by
\begin{equation}
    \tilde{\cal E}_{0}={r\over2}+\sum_{\rho\in\Delta_{+}}g_{|\rho|}.
    \label{defE0}
\end{equation}
which characterizes the ground state energy of the rational
Calogero-Moser model with harmonic confining force, see, for example,
(2.21) of \cite{kps}. This fact is closely related with
the $sl(2,{\bf R})$
algebra for rational Calogero-Moser models discussed by many authors,
see for example \cite{Pere1}-\cite{Br},\cite{Heck,Gonera1}.
We will not discuss $G_j^{(2)}$ any longer in this paper,
except for some comments in the final section.

 %%%%%%%%%%%%%%%%%%%%%%%%%%%%%%%%%%%%%%%%%%%%%
\section{Basic Commutation Relations}
\label{basecomm}
\setcounter{equation}{0}

Typical generators of the quadratic algebra are $\{F_j\}$'s
(\ref{jtotsum}) and
$\{H_{k,l}\}$'s (\ref{defHjk}).  Namely they are either linear in
$\{F_j\}$'s or bi-linear combinations of $\{F_j\}$'s and $\{G_k\}$'s.
As will be clear in later discussions, see for example (\ref{defhkls}),
the set of $\{F_j\}$'s must be understood in the broadest sense to
include the dependent ones. That is, any polynomials in the independent
$r$ involutive conserved quantities are allowed. For example, $F_j$ for
$j\neq 1+exponent$ or  $j>h$ (the Coxeter number) enter into
the theory naturally. Likewise, the set of $\{G_j\}$'s include
the dependent ones, which are independent ones times any
polynomial in $\{F_k\}$'s. In order to explore and present the
full content of the quadratic algebra, we need to evaluate the
commutators like:
\begin{equation}
[F_j,F_k],\quad [F_j,G_k],\quad [G_j,G_k].
\label{3coms}
\end{equation}
For this purpose the Dunkl operators \cite{Dunk} or $\ell$ operators
which are the {\em vector} version of the Lax matrix operator $L$
\cite{kps} are useful:
\begin{equation}
  \ell_\mu=\ell\!\cdot\!\mu=p\cdot\!\mu +
  i\sum_{\rho\in\Delta_+}g_{|\rho|}
\frac{\rho.\mu}{\rho.q}\,\check{s}_\rho,
  \qquad \mu\in{\mathcal R},
\label{ldef}
\end{equation}
in which another reflection operator $\check{s}_{\rho}$ acts on a (wave)
function $f$ of $q$ as
\begin{equation}
f\to \check{s}_{\rho}f: \quad  (\check{s}_{\rho}f)(q)=f(s_{\rho}(q)).
\label{scdef}
\end{equation}
The $\ell$ operator is  linear in \(\mu\),  Coxeter covariant
and hermitian:
\begin{equation}
   \check{s}_{\rho}\ell_{\mu}\check{s}_{\rho}=\ell_{s_{\rho}(\mu)},\quad
   \ell_{\mu}=\ell_{\mu}^\dagger, \quad \forall \rho\in\Delta.
\end{equation}
It is shown \cite{kps} that the Hilbert space of any quantum Calogero-Moser
system consists of Coxeter invariant wavefunctions. That is, they satisfy
\begin{equation}
\check{s}_{\rho}\psi=\psi,\quad \forall\rho\in\Delta.
\label{coxinv}
\end{equation}
It is well-known that the $\ell$ operators for the rational
Calogero-Moser models commute:
\begin{equation}
[\ell_\mu,\ell_\nu]=0,\quad \forall \mu,\nu\in{\cal R}.
\label{lcomm}
\end{equation}
The relationship between $L$ and $\ell$ is simple.
For any Coxeter invariant function $\psi$, $F_k\psi$
and $G_k\psi$ is Coxeter invariant too, and
we have \cite{kps}:
\begin{eqnarray}
 F_k\psi&\equiv& \mbox{Ts}(L^k)\psi\ \
=\sum_{\mu\in{\mathcal R}} \ell_\mu^k\psi,
 \ \ \,\qquad \forall k\in{\bf Z}_+,
\label{fkpsi}\\
 G_k\psi&\equiv& \mbox{Ts}(QL^k)\psi
=\sum_{\mu\in{\mathcal R}} q\cdot\!\mu\,\ell_\mu^k\psi,
 \quad \forall k\in{\bf Z}_+.
\label{gkpsi}
\end{eqnarray}
The involution of $\{F_j\}$'s  is a simple consequence of
(\ref{lcomm}) and (\ref{fkpsi}):
\begin{equation}
[F_j,F_k]=0,\quad \forall j,k\in {\bf Z}_+,
\label{fjk}
\end{equation}
which is a well-known result.

For the evaluation of the second and third types of commutators
in (\ref{3coms}) we need to know in general
\begin{equation}
[\ell_\mu^n\,, \,q\cdot\!\nu\,\ell_\nu^m].
\end{equation}
It is straightforward to show by induction
\begin{equation}
    [\ell^j_\mu,\,q\cdot\!\nu]=-i\left[j\,\mu\cdot\nu \,\ell_\mu^{j-1}
    +\sum_{\rho\in\Delta_+}g_{|\rho|}(\rho\cdot\mu)(\rho^\vee\!\cdot\nu)
     \frac{\ell_\mu^j-\ell^j_{s_\rho(\mu)}}
      {\ell_\mu-\ell_{s_{\rho}(\mu)}}\check{s}_\rho\right],
\label{ind}
\end{equation}
starting from
\begin{equation}
[\ell_\mu,\,q\cdot\!\nu]=-i\left[\mu\cdot\nu I
+\sum_{\rho\in\Delta_+}g_{|\rho|}(\rho\cdot\mu)
(\rho^\vee\!\cdot\nu)\check{s}_\rho\right],
\end{equation}
and
\begin{equation}
 [\ell^2_\mu,\,q\cdot\!\nu]=-i\left[2(\mu\cdot\nu)\, \ell_\mu
+\sum_{\rho\in\Delta_+}g_{|\rho|}(\rho\cdot\mu)(\rho^\vee\!\cdot\nu)
       \frac{\ell_\mu^2-\ell^2_{s_\rho(\mu)}}
     {\ell_\mu-\ell_{s_{\rho}(\mu)}}\check{s}_\rho\right].
\end{equation}
Here the fraction of operators,
$\ell_\mu^j-\ell^j_{s_\rho(\mu)}/{\ell_\mu-\ell_{s_{\rho}(\mu)}}$,
is well defined since the $\ell$ operators commute with each other,
(\ref{lcomm}). For example, we have
$\ell_\mu^2-\ell^2_{s_\rho(\mu)}/{\ell_\mu-\ell_{s_{\rho}(\mu)}}
=\ell_\mu+\ell_{s_{\rho}(\mu)}$.
Thus we arrive at
\begin{eqnarray}
   [\ell^j_\mu,\,q\cdot\!\nu\,\ell^k_\nu]&=
   &[\ell^j_\mu,\,q\cdot\!\nu]\ell^k_\nu\nonumber\\
   &=&-i\left[j\,\mu\cdot\nu \,\ell_\mu^{j-1}\ell^k_\nu
   +\sum_{\rho\in\Delta_+}g_{|\rho|}(\rho\cdot\mu)(\rho^\vee\!\cdot\nu)
   \frac{\ell_\mu^j-\ell^j_{s_\rho(\mu)}}
{\ell_\mu-\ell_{s_{\rho}(\mu)}}
\,\ell^k_{s_{\rho}(\nu)}\check{s}_\rho\right].
   \label{lnqlm}
\end{eqnarray}
The second term in
the right hand side of (\ref{lnqlm}) vanishes when summed over $\mu$:
\begin{equation}
   V\equiv\sum_{\mu\in{\mathcal R}}
   g_{|\rho|}(\rho\cdot\mu)(\rho^\vee\!\cdot\nu)
   \frac{\ell_\mu^j-\ell_{s_\rho(\mu)}^j}
   {\ell_\mu-\ell_{s_\rho(\mu)}}\,
   \ell^k_{s_\rho(\nu)}\check{s}_\rho=0.
   \label{vvani}
\end{equation}
This can be seen as follows.
The set ${\mathcal R}$ is Coxeter invariant, {\em i.e.,\/}
$s_\rho({\mathcal R})={\mathcal R}$.
Consider the change of variables
$\mu^\prime=s_\rho(\mu)$, then $\mu=s_\rho(\mu^\prime)$
and
\begin{eqnarray*}
V &=& \sum_{\mu^\prime\in{\mathcal R}}
g_{|\rho|}(\rho\cdot s_\rho(\mu^\prime))(\rho^\vee\!\cdot\nu)
\frac{\ell_{s_\rho(\mu^\prime)}^j-\ell_{\mu^\prime}^j}
{\ell_{s_\rho(\mu^\prime)}-
\ell_{\mu^\prime}}\,\ell^k_{s_\rho(\nu)}\check{s}_\rho\\
&=& \sum_{\mu^\prime\in{\mathcal R}}
g_{|\rho|}(-\rho\cdot\mu^\prime)(\rho^\vee\!\cdot\nu)
\frac{\ell_{\mu^\prime}^j-\ell_{s_\rho(\mu^\prime)}^j}
{\ell_{\mu^\prime}-\ell_{s_\rho(\mu^\prime)}}\,
\ell^k_{s_\rho(\nu)}\check{s}_\rho=-V.
\end{eqnarray*}
By summing over $\mu$ and $\nu$, we obtain from (\ref{lnqlm})
\begin{equation}
[\sum_{\mu\in{\cal R}}\ell_\mu^j,
\sum_{\nu\in{\cal R}}q\cdot\nu\,\ell_\nu^k]
=-ij\sum_{\mu,\nu\in{\cal R}}
(\mu\cdot\nu)\ell_\mu^{j-1}\ell^k_\nu.
\end{equation}
The right hand side is a Coxeter invariant polynomial in $\ell_\mu$,
which corresponds to a polynomial
in $\{F_j\}$ to be denoted by $F_{k,j}$:
\begin{equation}
\sum_{\mu,\nu\in{\cal R}}
(\mu\cdot\nu)\ell_\mu^{j-1}\ell^k_\nu\psi\equiv F_{k,j}\psi,
\qquad \psi:\ \mbox{Coxeter invariant}.
\label{fjkdef}
\end{equation}
Thus we arrive at
\begin{eqnarray}
 i[F_j,G_k]&=&jF_{k,j}
\label{fgcomm}\\
\ [F_n,F_{k,j}]&=&0,\quad \forall n\in{\bf Z}.
\label{fjkvan}
\end{eqnarray}

When the set of vectors ${\cal R}$ consists of orthonormal vectors,
for example, the vector representation of $A_r$ embedded in  an
$r+1$ dimensional space,
or vector representations of $C_r$ and $D_r$, or the set of short roots
of $B_r$, the above $F_{k,j}$ has a simpler expression.  In such cases,
only $\mu=\pm\nu$ terms in (\ref{fjkdef}) survive and we have
\begin{equation}
\sum_{\mu,\nu\in{\cal R}}
(\mu\cdot\nu)\ell_\mu^{j-1}\ell^k_\nu=
\left\{\begin{array}{c}
C_{\cal R}\sum_{\nu\in{\cal R}}
\ell_\nu^{j+k-1}\\[10pt]
0
\end{array}
\right. ,
\end{equation}
in which $C_{\cal R}$ is defined by (\ref{defcr}).
That is, (\ref{fgcomm}) is replaced by a more explicit formula
\begin{equation}
i[F_j,G_k]=jC_{\cal R}F_{j+k-1},
\label{fgsimp}
\end{equation}
which was reported in Kuznetsov's paper for $A_r$ case \cite{Kuz}
($C_{\cal R}=1$).
(In the above formula we assume that neither $F_j$ nor $G_k$ vanish.)
As for the extra exponent at $r-1$ in $D_r$ theory, the corresponding $F$
and
$G$ operators are best expressed by $\ell$ operators in the orthonormal
basis:
\begin{equation}
F_{r^\prime}\leftrightarrow \ell_1\cdots\ell_r,\qquad
G_{r^\prime-1}\leftrightarrow
\sum_{j=1}^r q_j\ell_1\cdots\tilde{\ell}_j\cdots\ell_r,
\end{equation}
in which $\tilde{\ell}_j$ means that the factor is missing.

\bigskip
The general commutation relations (\ref{fgcomm}), (\ref{fjkvan}) provide
the algebraic linearization
of the Hamiltonian systems generated by the higher
conserved quantities $\{F_j\}$.

\bigskip
{\bf Proposition \ref{basecomm}.1}

The Hamiltonian system generated by the higher conserved
quantity $F_j$ (\ref{jtotsum}) of quantum Calogero-Moser
system (\ref{ratHam}) is superintegrable for
any root system.

{\bf Proof.}
        On top of  the $D$ first integrals $F_{k}$, we have
the $D(D-1)/2$ extra first integrals for the Hamiltonian $F_j$:
\begin{eqnarray}
H_{k,k'}^{(j)}&=&F_{k,j}G_{k'}-F_{k',j}G_{k},\label{defHjkkp}\\
{d{H}_{k,k'}^{(j)}\over{dt_j}}&=
&i[F_j, H_{k,k'}^{(j)}]=0.\label{Hjkjcons}
\end{eqnarray}

%%%%%%%%%%%%%%%%%%%%%%%%%%%%%%%%%%%%%%%%%%%%%
\section{Quadratic Algebra}
\label{quadal}
\setcounter{equation}{0}

In order to evaluate the commutators among various $\{H_{k,k'}^{(j)}\}$'s
we need the knowledge of the third type of commutators in (\ref{3coms}),
that
is $[G_j,G_k]$.
From (\ref{lnqlm}) we have
 \begin{eqnarray}
&&[q\cdot\mu \,\ell_\mu^j,\ q\cdot\nu \,\ell_\nu^k]\nonumber\\
      &=&
    q\cdot\mu\,[\ell_\mu^j,\ q\cdot\nu \, \ell_\nu^k] + [q\cdot\mu,\
q\cdot\nu\,
    \ell_\nu^k]\,\ell_\mu^j\nonumber\\
    &=& -i\left\{(\mu\cdot\nu) j (q\cdot\mu)\, \ell_\mu^{j-1}\ell_\nu^k
    +
\sum_{\rho\in\Delta_+}g_{|\rho|}(q\cdot\mu)(\rho\cdot\mu)
    (\rho^{\vee}\cdot\nu)
    \frac{\ell_\mu^j-\ell^j_{s_\rho(\mu)}}
    {\ell_\mu-\ell_{s_\rho(\mu)}}\,\ell^k_{s_\rho(\nu)} \,
       \check{s}_\rho  \right\}\nonumber\\
   & & +i\left\{(\mu\cdot\nu) k (q\cdot\nu)\, \ell_\nu^{k-1}\ell_\mu^j
   +
\sum_{\rho\in\Delta_+}g_{|\rho|}(q\cdot\nu)(\rho\cdot\nu)
(\rho^{\vee}\cdot\mu)
         \frac{\ell_\nu^k-l^k_{s_\rho(\nu)}}
{\ell_\nu-\ell_{s_\rho(\nu)}}\,\ell^j_{s_\rho(\mu)} \,
    \check{s}_\rho\right\}. \label{gjgkcomm}
\end{eqnarray}
As in the previous case (\ref{vvani}), the coupling constant dependent
terms, that is
the second and fourth terms in (\ref{gjgkcomm}) cancel with each other
when summed over $\mu$ and $\nu$:
\begin{eqnarray}
   &&\left[\sum_{\mu\in{\mathcal R}}(q\cdot\mu)\,\ell_\mu^j\, ,\
       \sum_{\nu\in{\mathcal R}}
           (q\cdot\nu)\,\ell_\nu^k\right]\nonumber\\
   &&\hspace*{2cm}  = -i\sum_{\mu,\nu\in{\mathcal R}}
   \left\{j\, (\nu\cdot\mu) (q\cdot\mu)
   \ell_\mu^{j-1}\ell_\nu^{k}- k\, (\mu\cdot\nu)  (q\cdot\nu)
   \ell_\nu^{k-1}\ell_\mu^{j}\right\}.
\label{gjgkcomm2}
\end{eqnarray}
Both terms in the right hand side are Coxeter invariant polynomials
in $q$ and $\ell$ which are linear in $q$ and of degree $j+k-1$ in $\ell$.
Therefore they are expressible as linear combination of
$\{G_l\}$'s or polynomials in $\{F_m\}$'s multiplied on them.
This can be checked by direct calculation or by using the Jacobi identity
on the left hand side.
Thus we express
\begin{equation}
\begin{array}{r}
\sum_{\mu,\nu\in{\cal R}}
(\mu\cdot\nu)(q\cdot\mu)
   \ell_\mu^{j-1}\,\ell_\nu^{k}\,\psi\equiv G_{j,k}\,\psi,\\[10pt]
\sum_{\mu,\nu\in{\cal R}}
(\mu\cdot\nu)  (q\cdot\nu)
   \ell_\mu^{j}\,\ell_\nu^{k-1}\psi\equiv G_{k,j}\,\psi,
\end{array}
\quad \psi:\ \mbox{Coxeter invariant}.
\label{gjkdef}
\end{equation}
We arrive at the following general commutation relation
\begin{equation}
i\,[G_j,G_k]=j\,G_{j,k}-k\,G_{k,j}.
\end{equation}
These $\{G_{j,k}\}$'s satisfy the same type of commutation relations as
above.

When the set of vectors ${\cal R}$ consists of orthonormal vectors,
we have
\begin{equation}
\sum_{\mu,\nu\in{\cal R}}
(\mu\cdot\nu)(q\cdot\mu)\ell_\mu^{j-1}\ell^k_\nu=
\sum_{\mu,\nu\in{\cal R}}
(\mu\cdot\nu)  (q\cdot\nu)
   \ell_\mu^{j}\,\ell_\nu^{k-1}=
\left\{\begin{array}{c}
C_{\cal R}\sum_{\nu\in{\cal R}}(q\cdot\nu)\,
\ell_\nu^{j+k-1}\\[10pt]
0
\end{array}
\right. .
\end{equation}
This leads to a simplified commutation relation
\begin{equation}
i\,[G_j,G_k]=(j-k)C_{\cal R}\,G_{j+k-1},
\label{ggsimp}
\end{equation}
which was reported in Kuznetsov's paper for $A_r$ case \cite{Kuz}
($C_{\cal R}=1$).

\bigskip
To sum up, we have obtained the following general commutation
relations:
\begin{eqnarray}
[F_j,F_k]&=&0,\\
i[F_j,G_k]&=&j\,F_{k,j},\label{fgf}\\
i[G_j,G_k]&=&j\,G_{j,k}-k\,G_{k,j}.
\end{eqnarray}
By using these the operators $\{F_j\}$'s and $\{H_{k,l}^{(m)}\}$'s defined
by
\begin{equation}
H_{k,l}^{(m)}=F_{k,m}G_{l}-F_{l,m}G_{k},\quad
H_{k,l}^{(m)}=-H_{l,k}^{(m)},
\label{defhkls}
\end{equation}
generate a quadratic algebra
\begin{eqnarray}
[F_j,F_k]&=&0,\\
i[F_j,H_{k,l}^{(m)}]&=&j\,(F_{k,m}F_{l,j}-F_{l,m}F_{k,j}),\\
i[H_{k,l}^{(m)},H_{k',l'}^{(m')}]&=&\mbox{quadratic in}\ H_{r,s}^{(n)}\
\mbox{and}\ F_t.
\label{hhcom}
\end{eqnarray}
This is the quadratic algebra of the quantum rational Calogero-Moser models
based on any root systems. For the classical root systems it can
be simplified by using the relations (\ref{fgsimp}) and (\ref{ggsimp})
to the forms given in Kuznetsov's paper \cite{Kuz}.
It characterizes the superintegrability structure of quantum models.
In  applications for specific models, the indices of $\{F\}$'s and $\{G\}$'s
and $\{H\}$'s must be chosen properly. This would give more specific forms
of the quadratic algebra relations.

%%%%%%%%%%%%%%%%%%%%%%%%%%%%%%%%%%%%%%%%%%%%%
\section{Rational Potential Model with Harmonic Confining Force}
\label{harm}
\setcounter{equation}{0}

The arguments for the algebraic linearization for the {\em quantum}
rational potential model with harmonic confining force go almost parallel
with the classical ones. So we present only the key formulas.
We have to note the coupling dependence is changed
from $g_{|\rho|}^2$ (classical) to $g_{|\rho|}(g_{|\rho|}-1)$ (quantum)
and instead of {\em trace} (Tr, classical)
we need the {\em total sum} (Ts, quantum).
The Hamiltonian is now:
\begin{equation}
    {\cal H}_{\omega}={1\over2}p^2+{1\over2}\omega^2q^2+
       {1\over2}\sum_{\rho\in\Delta_+}
     {g_{|\rho|}(g_{|\rho|}-1) |\rho|^{2}\over{(\rho\cdot
          q)^2}}.
     \label{harmham}
\end{equation}
With the same matrix operators $L$, $M$ and $Q$ as in the rational case
the equations of motion can be expressed in matrix forms:
\begin{equation}\label{RCPL}
  \dot{L}=[L,M]-\omega^2Q, \qquad
  \dot{Q}=[Q,M]+L.
\end{equation}
Introduce the matrices
\begin{equation}
L^{\pm}=L\pm i\omega Q
\end{equation}
whose time evolution read
\begin{equation}
\dot{L}^\pm
 = [L^\pm, M]\pm i\omega L^\pm.
\end{equation}
They can be cast into a Lax form for ${\mathcal L}=L^+L^-$ as
\begin{equation}
    \dot{{\cal L}}=[{\cal L},M].
\label{calLlax}
\end{equation}
Consider then the functions:
\begin{equation}
F_{k}=\mbox{Ts}(L^{+}{\cal L}^{k}), \qquad
G_{k}=\mbox{Ts}(L^{-}{\cal L}^{k}). \label{fgharm}
\end{equation}
The time evolution yields
\begin{equation}
\dot{F}_{k}=i\omega F_{k}, \qquad
\dot{G}_{k}=-i\omega G_{k}.
\end{equation}
Thus these functions provide the algebraic
linearization of the quantum system.

%%%%%%%%%%%%%%%%%%%%%%%%%%%%%%
\section{Rational Model with a
Quartic Potential}
\label{quartic}
\setcounter{equation}{0}

As proved by Fran\c{c}oise and Ragnisco \cite{fr} for the models based on
the $A$ type root systems  and by us \cite{cfs} for the models
based on any root systems,
the rational Calogero-Moser model can be deformed into an
integrable system by adding a quartic potential at the {\em classical}
level.
Here we provide a partial result at the quantum level.
The equation of motion can be cast into  Lax type equations but
they fail to produce  conserved quantities.

Define again the same matrices $L$, $Q$, $X$ and $M$. Let
\begin{equation}
h(Q)=aQ+bQ^{2}
\end{equation}
be a matrix quadratic in $Q$; $(a,b)$ are just two new independent
parameters.
The perturbed Hamiltonian is now:
\begin{equation}
    {\cal H}_{h}\thinspace {\propto}
\thinspace \mbox{Ts}(L^{2}+h(Q)^{2}).
    \label{quartham}
\end{equation}
Like in the classical theory, the equations of motion can be cast into
matrix forms by defining
\begin{equation}
 L^{\pm}=L\, {\pm} \, {\rm i} h(Q),\quad
 {\cal L}_1=L^{+}L^{-},\quad {\cal L}_2=L^{-}L^{+}.
\end{equation}
Though care is needed for the quantum non-commutativity, the calculation is
essentially the same as in the classical theory and we arrive at
the time evolution of $L^\pm$ and ${\cal L}_1$,  ${\cal L}_2$:
\begin{equation}
\dot{L}^{\pm}=[L^{\pm},\, M] \pm
i{1\over2}\left(h'(Q)L^{\pm}+L^{\pm}h'(Q)\right),
\end{equation}
\begin{equation}
\dot{\cal L}_1=[{\cal L}_1,\, M-{i\over2}h'(Q)],\quad
\dot{\cal L}_2=[{\cal L}_2,\, M+{i\over2}h'(Q)].
\end{equation}
Because of the added term $\pm {i\over2}h'(Q)$ to the $M$ matrix,
it loses the sum up to zero property (\ref{zerosum}) and thus {\em neither
trace nor total sum} of the powers of ${\cal L}_{1,2}$ are conserved
at the quantum level.

%%%%%%%%%%%%%%%%%%%%%%%%%%%%%%
\section{Trigonometric  Calogero-Sutherland
Model}
\label{trig}
\setcounter{equation}{0}

The algebraic linearization of the trigonometric (hyperbolic)
{\em classical} Calogero-Sutherland
model was shown in our previous paper \cite{cfs} for the root systems
which have minimal representations,
that is $A$ and $D$ series and $E_6$ and $E_7$. These are all
simply laced algebras and all the roots have the same length.
The quantum Hamiltonian reads
\begin{equation}
    {\cal H}={1\over2}p^2+
       {1\over2}g(g-1)|\alpha|^{2}\sum_{\alpha\in\Delta_+}
          {1\over{\sin^{2}(\alpha\cdot q)}}.
          \label{(5.1)}
\end{equation}
        We consider the Lax matrices:
\begin{eqnarray}
L&=&p\cdot\hat{H}+X, \qquad
X=ig\sum_{\rho\in\Delta_{+}}\thinspace
   (\rho\cdot\hat{H})\thinspace{1\over{\sin(\rho\cdot
   q)}}\thinspace\hat{s}_{\rho}, \label{(5.2b)}\\
M&=&-{ig|\rho|^2\over2}\sum_{\rho\in\Delta_{+}}\thinspace
   {\cos(\rho\cdot q)\over{\sin^{2}(\rho\cdot q)}}
   \thinspace(\hat{s}_{\rho}-I),
\label{(5.2c)}
\end{eqnarray}
and diagonal matrices:
\begin{equation}
R={{\rm e}}^{2iQ},\qquad
Q=q\cdot\hat{H}.
\end{equation}
Then, as in the classical case, we obtain
\begin{equation}
\dot{L}=[L,M]
\end{equation} and
\begin{equation}
\dot{R}=[R,M]+i(RL+LR).
\end{equation}
This is because the main formula of proof
in the classical theory, eq(5.8) in
\cite{cfs} is the same in quantum theory.
We only have to change the definition of $a_k$ eq(5.4a) and $b_k$ eq(5.4b)
in
\cite{cfs} in order to accommodate for the quantum non-commutativity.
Define
\begin{equation}\label{TFks}
  a_j=\mbox{Ts}(L^j),\quad
  b_j=\mbox{Ts}\sum_{k=0}^j{j \choose k}
    \, L^k R\, L^{j-k},
\end{equation}
then we obtain
\begin{equation}
\dot{a}_j=0,\quad \dot{b}_j=ib_{j+1}.
\end{equation}
This provides the algebraic linearization of quantum
trigonometric (hyperbolic) Calogero-Sutherland for the
$A$ and $D$ series and $E_6$ and $E_7$ root systems.
The models with hyperbolic potential can be discussed in a
similar way.
See also \cite{Gonera2,ttw} in this connection.

%%%%%%%%%%%%%%%%%%%%%%%%%%%%%%
\section{Comments on the Hermiticity of Algebra Generators}
\label{comm}
\setcounter{equation}{0}
In quantum mechanics physical quantities or the observables are
described by hermitian operators in Hilbert space \cite{Gonera1}.
The hermiticity of $F_j$, (\ref{jtotsum}) is obvious from that of $L$.
The original definition of $G_j$, (\ref{Qdef}) is not hermitian.
With the following redefinition of hermitian $G_j$,
\begin{equation}
G_j=\mbox{Ts}\sum_{k=0}^j(L^kQL^{j-k})/(j+1),
\end{equation}
it satisfies the same formula (\ref{Fgrel}).
Whereas the definition of $F_{k,j}$ (\ref{fjkdef}) remains the same,
that of  $H_{k,k'}^{(j)}$ (\ref{defHjkkp}) should be changed to
a hermitian form
\begin{equation}
2H_{k,k'}^{(j)}=F_{k,j}G_{k'}+G_{k'}F_{k,j}-
F_{k',j}G_{k}-G_{k}F_{k',j}.\label{newdefHjkkp}
\end{equation}

A formulation with  explicitly hermitian $G_j$ could have been achieved by
\begin{equation}
G_{j-1}\propto i[q^2,\,F_j],
\end{equation}
which is closely related with the extension of the $sl(2,{\bf R})$ algebra
\cite{Pere1}-\cite{Br},\cite{UjWa,Heck,Gonera1}. This also explains the
assertion that independent $\{G_j\}$'s appear at $j=exponent$.

We chose the current presentation in order to avoid excessively
complicated looking formulas and to allow
an easy comparison with the original work \cite{Kuz} on the quadratic
algebra.

\section*{Acknowledgements}
\setcounter{equation}{0}
R. C. and R. S. thank the Universit\'e P.-M. Curie, Paris VI
for hospitality.
        J.-P.F. is partially supported by a grant from the
French Ministry of Education and Research to the Laboratory
``GSIB", Universit\'e P.-M. Curie, Paris VI.
        R. S. is partially supported  by the Grant-in-aid from the
Ministry of Education, Culture, Sports, Science and Technology, Japan,
priority area (\#707) ``Supersymmetry and unified theory of elementary
particles".
%%%%%%%%%%%%%%%%%%%%%%%%%%%%%%%%%%%%%%%%%%%%%%%%%%%%%


\begin{thebibliography}{99}
\bibitem{Cal}  F.\, Calogero, ``Solution of the one-dimensional
\(N\)-body problem with quadratic and/or inversely quadratic pair
potentials", J. Math. Phys. {\bf 12} (1971) 419-436.
\bibitem{Sut}
B.\, Sutherland, ``Exact results for a quantum many-body problem in
one-dimension. II'', Phys. Rev. {\bf A5} (1972) 1372-1376.
\bibitem{CalMo}
J.\, Moser, ``Three integrable Hamiltonian systems connected with
isospectral deformations'',  Adv. Math. {\bf 16} (1975) 197-220;\
J.\, Moser,  ``Integrable systems of non-linear evolution equations",
in {\it Dynamical Systems, Theory and Applications\/};\
J. Moser, ed., Lecture Notes in Physics {\bf 38} (1975),
Springer-Verlag;\
F.\,Calogero, C.\, Marchioro and O.\, Ragnisco, ``Exact solution of the
classical and quantal one-dimensional many body problems with
the two body potential \(V_{a}(x)=g^2a^2/\sinh^2\,ax\)'', Lett. Nuovo
Cim. {\bf 13} (1975) 383-387;\
F.\,Calogero,``Exactly solvable one-dimensional many body problems'',
Lett. Nuovo Cim. {\bf 13} (1975) 411-416.

\bibitem{Woj}
S.\, Wojciechowski, ``Involutive set of integrals for completely
integrable many-body problems with pair interaction",
Lett. Nuouv. Cim. {\bf 18} (1976) 103-107;
%%CITATION = NCLTA,18,103;%%
``Superintegrability of the Calogero-Moser system'',
Phys. Lett. {\bf A95} (1983) 279-281.


%\cite{Kuznetsov:1995xn}
\bibitem{Kuz}
V.~B.~Kuznetsov,
``Hidden symmetry of the quantum Calogero-Moser system'',
Phys. Lett. {\bf A218} (1996) 212-222, {\tt solv-int/9509001}.
%%CITATION = SOLV-INT 9509001;%%

\bibitem{UjWa}
 H.\, Ujino,  M.\, Wadati and K. Hikami, ``The quantum Calogero-Moser
model: algebraic structures", J. Phys. Soc. Jpn.
 {\bf 62} (1993) 3035--3043;



\bibitem{cf}{R.\,Caseiro and J.-P.\, Fran\c{c}oise,}{`` Algebraically
linearizable dynamical systems'', to appear.}

\bibitem{cfs}
R.\,Caseiro, J.-P.\,Fran\c{c}oise and R.\,Sasaki,
``Algebraic Linearization of Dynamics of Calogero Type for any
Coxeter Group'',
J.\ Math.\ Phys.\ {\bf 41} (2000) 4679-4986,
{\tt hep-th/0001074}.
%%CITATION = HEP-TH 0001074;%%


\bibitem{Zam}
A.~B.~Zamolodchikov,
``Infinite Additional Symmetries In Two-Dimensional
Conformal Quantum Field
Theory,''
Theor.\ Math.\ Phys.\ {\bf 65} (1985) 1205-1213.
%%CITATION = TMPHA,65,1205;%%


\bibitem{Freidel-Maill}
L.~Freidel and J.~M.~Maillet,
``Quadratic algebras and integrable systems,''
Phys.\ Lett.\ {\bf B262} (1991) 278-284.
%%CITATION = PHLTA,B262,278;%%


\bibitem{OP1} M.\,A.\, Olshanetsky and A.\,M.\, Perelomov,
``Completely integrable Hamiltonian systems connected with
 semisimple Lie algebras",
 Inventions Math. {\bf 37} (1976), 93-108;
 ``Classical integrable finite-dimensional systems related to Lie
 algebras'',
 Phys. Rep.  {\bf C71} (1981), 314-400.

 \bibitem{bcs2}  A.\,J.\, Bordner, E.\, Corrigan and R.\, Sasaki,
``Generalized Calogero-Moser models and  universal Lax pair operators'',
 Prog. Theor. Phys. {\bf 102}  (1999)  499-529,
 {\tt  hep-th/9905011}.
%%CITATION = HEP-TH 9905011;%%

 \bibitem{DHoker_Phong}
E.\,D'Hoker and D.\,H.\,Phong, ``Calogero-Moser
Lax pairs with spectral parameter for general Lie algebras'',
Nucl. Phys. {\bf B530} (1998) 537-610, {\tt hep-th/9804124}.
%%CITATION = HEP-TH 9804124;%%




\bibitem{bcs1}
 A.\,J.\, Bordner, E.\, Corrigan and R.\, Sasaki,
``Calogero-Moser models I: a new formulation'',
Prog. Theor. Phys. {\bf 100} (1998) 1107-1129, {\tt hep-th/9805106};
%%CITATION = HEP-TH 9805106;%%
A.\,J.\, Bordner,   R.\,
Sasaki and K.\, Takasaki, ``Calogero-Moser models II:
symmetries and foldings'', Prog. Theor. Phys. {\bf
101} (1999) 487-518, {\tt hep-th/9809068};
%%CITATION = HEP-TH 9809068;%%
A.\,J.\,Bordner and R.\,Sasaki, ``Calogero-Moser models III: elliptic
potentials and
twisting'', Prog. Theor. Phys. {\bf 101} (1999) 799-829, {\tt
hep-th/9812232};
%%CITATION = HEP-TH 9812232;%%
S.\,P.\, Khastgir, R.\, Sasaki and K.\, Takasaki,
``Calogero-Moser models IV: Limits to Toda theory",
 Prog. Theor. Phys. {\bf 102}  (1999), 749-776, {\tt
hep-th/9907102}.
%%CITATION = HEP-TH 9907102;%%


\bibitem{bms}  A.\,J.\, Bordner, N.\,S.\, Manton and R.\, Sasaki,
``Calogero-Moser models V:  Supersymmetry,
and Quantum Lax Pair", Prog. Theor. Phys. {\bf 103} (2000) 463-487,
{\tt hep-th/9910033}.
%%CITATION = HEP-TH 9910033;%%

\bibitem{Dunk}
C.\,F.\,Dunkl, ``Differential-difference operators associated to
reflection groups", Trans. Amer. Math. Soc. {\bf 311} (1989) 167-183.

\bibitem{Heck}
G.\,J.\, Heckman, ``A remark on the Dunkl differential-difference
operators", in W.\, Barker and P.\, Sally (eds.) ``Harmonic analysis
on reductive groups", Birkh\"auser, Basel (1991).

\bibitem{kps}
S.\, P.\, Khastgir, A.\, J.\, Pocklington and R.\, Sasaki,
``Quantum Calogero-Moser Models: Integrability for all Root Systems'',
J.\ Phys. {\bf A33} (2000) 9033-9064,
{\tt hep-th/0005277}.
%%CITATION = HEP-TH 0005277;%%



\bibitem{ks2}
S.\,P.\, Khastgir and R.\, Sasaki,
``Liouville Integrability of Classical Calogero-Moser Models'',
Phys. Lett. {\bf A279} (2001) 189-193,
{\tt hep-th/0005278}.
%%CITATION = HEP-TH 0005278;%%

\bibitem{Gonera1}
C.\,Gonera,
``A note on the superintegrability of quantum Calogero model'',
 Phys. Lett. {\bf A237} (1998) 365-368.


\bibitem{Ranada}
M.\,F.\, Ra\~{n}ada,
``Superintegrability of the Calogero-Moser system:
Constants of motion, master symmetries and time-dependent symmetries'',
J. Math. Phys. {\bf 40} (1999) 236-247.


\bibitem{Pere1}
A.\,M.\, Perelomov, ``Algebraic approach to the solution of a
one-dimensional model of interacting particles",
Theor. Math. Phys. {\bf 6} 263-282, (1971).

\bibitem{Gamb}
P.\,J.\, Gambardella, ``Exact results in quantum many-body systems of
interacting particles in many dimensions with \(\overline{SU(1,1)}\) as the
dynamical group",
J. Math. Phys. {\bf 16} 1172-1187 (1975).

\bibitem{Br}
L.\, Brink, T.\,H.\, Hansson and M.\,A.\, Vasiliev, `` Explicit solution
to
the \(N\) body Calogero problem", Phys. Lett. {\bf B286} (1992) 109-111,
{\tt hep-th/9206049};
L.\, Brink, T.\,H.\, Hansson, S.\, Konstein and M.\,A.\, Vasiliev,
``The Calogero model: anyonic representation, fermionic extension and
supersymmetry", Nucl. Phys. {\bf B401} (1993) 591-612,
{\tt hep-th/9302023};
L.\, Brink, A.\, Turbiner and N.\, Wyllard ``Hidden algebras of the
(super) Calogero and Sutherland models", J. Math. Phys. {\bf 39} (1998)
1285-1315,  {\tt hep-th/9705219}.


\bibitem{fr}{J.-P. Fran\c{c}oise and  O. Ragnisco,}{ }{``Matrix differential
equations and Hamiltonian systems of quartic type", Ann. Inst. H.
Poincar\'{e} {\bf 49} (1989) 369-375.}

\bibitem{Gonera2}
C.\,Gonera,
``On the superintegrability of Calogero-Moser-Sutherland model'',
J. Phys. {\bf A31} (1998) 4465-4472.

\bibitem{ttw}
P.\,Tempesta, A.\,Turbiner and P.\, Winternitz,
``Exact solvability of superintegrable systems",
{\tt hep-th/0011209}.

\end{thebibliography}
\end{document}